\documentclass[
reprint,
superscriptaddress,
frontmatterverbose, 
showpacs,
preprintnumbers,
nofootinbib,
amsmath,
amssymb,
aps,
floatfix,
twocolumn
]{revtex4-2}

\usepackage{graphicx}
\usepackage{dcolumn}
\usepackage{bm}
\usepackage{hyperref}
\usepackage{xcolor}

\begin{document}


\title{
Vortex Creep Heating vs. Dark Matter Heating in Neutron Stars}

\author{Motoko Fujiwara}
 \email{motoko.fujiwara@tum.de}
 \affiliation{Physik-Department, Technische Universit\"at, M\"unchen, James-Franck-Stra\ss e, 85748 Garching, Germany
}

\author{Koichi~Hamaguchi}
 \email{hama@hep-th.phys.s.u-tokyo.ac.jp}
 \affiliation{Kavli IPMU (WPI), University of Tokyo, Kashiwa, Chiba 277--8583, Japan
}
 \affiliation{Department of Physics, University of Tokyo, Bunkyo-ku, Tokyo
 113--0033, Japan
 }
\author{Natsumi~Nagata}
 \email{natsumi@hep-th.phys.s.u-tokyo.ac.jp}
  \affiliation{Department of Physics, University of Tokyo, Bunkyo-ku, Tokyo
 113--0033, Japan
 }
\author{Maura E. Ramirez-Quezada}
  \email{me.quezada@hep-th.phys.s.u-tokyo.ac.jp} 
   \affiliation{Department of Physics, University of Tokyo, Bunkyo-ku, Tokyo 113--0033, Japan
   }
    \affiliation{Dual CP Institute of High Energy Physics, C.P. 28045, Colima, M\'exico}

\date{\today}

\begin{abstract}
  Dark matter particles captured in neutron stars deposit their energy as heat. This DM heating effect can be observed only if it dominates over other internal heating effects in NSs. In this work, as an example of such an internal heating source, we consider the frictional heating caused by the creep motion of neutron superfluid vortex lines in the NS crust. The luminosity of this heating effect is controlled by the strength of the interaction between the vortex lines and nuclei in the crust, which can be estimated from the many-body calculation of a high-density nuclear system as well as through the temperature observation of old NSs. We show that both the temperature observation and theoretical calculation suggest that the vortex creep heating dominates over the DM heating. The vortex-nuclei interaction must be smaller than the estimated values by several orders of magnitude to overturn this. 

\end{abstract}

\maketitle


\section{Introduction}

Dark Matter (DM) is a mysterious gravitational source that remains one of the biggest enigmas in the universe. 
Among the potential candidates,  Weakly Interacting Massive Particles (WIMPs) are particularly intriguing. 
The thermal history of WIMPs in the expanding universe offers a natural explanation for the observed energy density of DM, as determined by observations of the Cosmic Microwave Background (CMB)~\cite{Planck:2018vyg}. 
WIMPs interact with particles in the Standard Model (SM), providing avenues to probe their properties. The different approaches, such as direct detection, indirect detection, and collider searches, provide valuable information on the properties of DM. However, despite extensive efforts, the conclusive identification of DM particles remains an ongoing challenge in the field of particle physics.

Recently, there has been growing interest in using Neutron Star (NS) temperature observation as a means of searching for WIMPs~\cite{Kouvaris:2007ay, Bertone:2007ae, Kouvaris:2010vv, deLavallaz:2010wp, Bramante:2017xlb, Baryakhtar:2017dbj, Raj:2017wrv,  Chen:2018ohx, Bell:2018pkk, Garani:2018kkd, Camargo:2019wou, Bell:2019pyc, Hamaguchi:2019oev, Garani:2019fpa, Acevedo:2019agu, Joglekar:2019vzy, Keung:2020teb,  Yanagi:2019zne, Joglekar:2020liw, Bell:2020jou, Dasgupta:2020dik, Bell:2020lmm, Anzuini:2021lnv, Zeng:2021moz, Bramante:2021dyx, Tinyakov:2021lnt,Maity:2021fxw, Fujiwara:2022uiq, Hamaguchi:2022wpz, Chatterjee:2022dhp, Coffey:2022eav, Acuna:2022ouv, Alvarez:2023fjj, Bramante:2023djs}.
By observing old NSs, we can explore the unknown properties of DM through its interactions with stellar matter. 
NSs possess a strong gravitational potential, causing  WIMP  particles to fall and become gravitationally bound to them.  
During this process, WIMP particles scatter off the stellar matter, losing some of their kinetic energy to the NS~\cite{Goldman:1989nd}. 
The capture of DM can occur via interactions with nucleons, leptons, and even hyperons.
If WIMPs lose enough energy, they will be captured within a NS core, where they thermalize and eventually annihilate into SM particles. 
The capture, thermalization, and annihilation process inject energy into the NS, serving as a late-time heating source of old isolated NSs~\cite{Kouvaris:2007ay}.
The heating effect typically predicts a surface temperature of a few thousand Kelvin~\cite{Chatterjee:2022dhp}, which is anomalously hotter compared to the prediction from the standard cooling theory for NSs~\cite{Yakovlev:1999sk,Yakovlev:2000jp,Yakovlev:2004iq, Page:2004fy, 2009ApJ...707.1131P, Potekhin:2015qsa} that are older than $10^6~\mathrm{yrs}$.  Therefore, by observing NSs colder than this predicted WIMP temperature, we may be able to place constraints on the WIMP-nucleon cross section.

In the context of NS temperature observations to probe DM properties, it is commonly assumed that the heating effects from DM dominate a late heating source for the star. However, observations have revealed the existence of old but still warm NSs with surface temperatures typically exceeding $10^5~\mathrm{K}$~\cite{Kargaltsev:2003eb, Mignani:2008jr, Durant:2011je, Rangelov:2016syg, Pavlov:2017eeu, Abramkin:2021tha, Abramkin:2021fzy}. The presence of such high temperatures in these compact objects suggests that additional internal heating mechanisms must be at play. Since DM heating alone cannot explain the observed temperatures, other heating sources must contribute to the thermal energy of these objects. For a review of such a heating mechanism in NSs, see Refs.~\cite{Schaab:1999as, Gonzalez:2010ta, Kopp:2022lru}. If NSs are primarily heated by internal mechanisms that predict a universal temperature as high as $10^5~\mathrm{K}$, as observed recently, it would indeed pose challenges for probing WIMP through future temperature observations~\cite{Hamaguchi:2019oev}. 

\textit{The vortex creep heating}~\cite{1984ApJ...276..325A, 1984ApJ...278..791A, 1985ApJ...288..191A, 1989ApJ...346..808S, 1989ApJ...346..823A, 1991ApJ...381L..47V, 1993ApJ...408..186U, 1995ApJ...448..294V, Schaab:1999as, Larson:1998it, Gonzalez:2010ta, Fujiwara:2023tmr} is a universal and internal heating mechanism. This heating is caused by friction associated with the creep motion of vortex lines in the neutron superfluid of the inner crust region. Its heating luminosity is proportional to the time derivative of the angular velocity of the NS, and the proportional coefficients (denoted by $J$ in this paper) are expected to be universal and determined by the nuclear-vortex interaction in the inner crust region. We can in principle evaluate this parameter $J$ by means of a nuclear many-body calculation~\cite{Avogadro:2006ed,Avogadro:2007kxn, Avogadro:2008uy,Klausner:2023ggr, 1988ApJ...328..680E, Pizzochero:1997iq,Donati:2003zz,Donati:2006mfa,Donati:2004gnw,Seveso:2016, Link:2008aq, Antonelli:2020umo, Bulgac:2013nmn, Wlazlowski:2016yoe, Link:2022vum}. It is also possible to determine this from the NS surface temperature observations by fitting the data with the prediction of the vortex creep heating mechanism. In Ref.~\cite{Fujiwara:2023tmr}, we showed that the values of $J$ obtained with these two approaches are consistent with each other. In the present paper, we discuss the implications of this result for the WIMP DM search through the NS surface temperature observations.

\section{Thermal history of neutron stars}
\label{sec:Thermal_history_of_NS}

We begin by reviewing the thermal history of NSs~\cite{Yakovlev:1999sk, Yakovlev:2000jp, Yakovlev:2004iq}.
The thermalisation process of  NSs is typically completed by $t \lesssim 10^{2}~\mathrm{yrs}$~\cite{1994ApJ...425..802L, Gnedin:2000me}, and the subsequent thermal evolution remains largely independent of the initial conditions for each star. The following equation describes the thermal evolution of NSs,
\begin{align}
  \mathcal{C}^\infty(T^\infty)  \frac{dT^\infty}{dt}  
  &=  
  -  L^\infty_\nu-  L^\infty_\gamma  +  L^\infty_{\rm  heat},
\end{align}
where $\mathcal{C}^\infty(T^\infty)$ is the heat capacity of the star. 
On the right-hand side, the terms with a negative sign, $L_\nu$ and $L_\gamma$, are redshifted luminosity for cooling by the neutrino and photon emissions, respectively. 
The term with a positive sign, $L^\infty_{\rm  heat}$, denotes the redshifted luminosity for heating if any exists.

The early phase of cooling for $t  \lesssim  10^5~\mathrm{yrs}$ occurs with neutrino emission such as direct Urca~\cite{Gamow:1941gis}, modified Urca~\cite{1995A&A...297..717Y,Gusakov:2002hh}, and Pair-breaking and formation (PBF) process~\cite{1976ApJ...205..541F,Voskresensky:1987hm,Senatorov:1987aa}. 
The late phase of cooling for $t  \gtrsim  10^5~\mathrm{yrs}$ is associated with photon emission, and the cooling luminosity is given by the Stefan-Boltzmann law. 
If NSs have a heating source, the associated heating luminosity and the photon cooling luminosity will eventually reach thermal equilibrium.
Therefore, the late-time surface temperature of NSs will be characterised by their luminosity through the following relation. 
\begin{align}
  L^\infty_{\rm  heat}  \simeq  L^\infty_{\gamma}  =  B(R_{\rm  NS})  \times  4  \pi  R_{\rm  NS}^2  \sigma_{\rm  SB}  T_{\rm  s}^4,
  \label{eq:L_heat=L_gamma}
\end{align}
where $R_{\rm  NS}$ is the NS radius, $T_{\rm  s}$ is the surface temperature, and $\sigma_{\rm  SB}$ is the Stefan-Boltzmann constant. 
The factor of $B(r)$, the time component of the Schwartzschild metric, represents the redshift effect on luminosity. 
In the following, we will consider two types of heating sources: DM heating and vortex creep heating.

\section{Dark matter heating}
\label{sec:DM_heating}

Dark matter particles accrete in NS cores and scatter with the stellar matter. In particular, WIMPs with a mass of $1~\mathrm{GeV}  \lesssim  m_{\chi}  \lesssim  1~\mathrm{PeV}$ will be efficiently captured gravitationally~\cite{Baryakhtar:2017dbj}.
For neutron targets, the threshold value of DM-neutron cross section for capture is derived as $\sigma_{\chi 
 n}  \simeq  1.7  \times  10^{-45}~\mathrm{cm}^2 ~(1.4  \times  10^{-44}~\mathrm{cm}^2)$  in Ref.~\cite{Bell:2020jou} (Ref.~\cite{Anzuini:2021lnv}).

The DM captured within a NS core will be thermalized and eventually annihilated into  SM particles. The capture, thermalization, and annihilation processes will inject energy into the NS and increase its temperature~\cite{Kouvaris:2007ay}. The general expression for the capture rate by nucleons and leptons in a NS is given by~\cite{Goldman:1989nd,Kouvaris:2007ay},
\begin{equation}
    C \! =  \!
    \frac{4\pi}{v_{\star}}
    \frac{\rho_\chi}{m_\chi}
    {\rm Erf} \! \left(\!\!\sqrt{\frac{3}{2}}\frac{v_{\star}}{v_d}\right)
    \!\!\int_0^{R_{\rm NS}}\!\!\!\! dr\frac{1-B(r)}{B(r)}  r^2  \Omega^-(r)  \eta(r)~,
\end{equation}
where $v_\star$ and $v_d$ are the NS and dispersion velocities, respectively. 
Here $\rho_\chi$ is the DM density which, considering local NSs, takes the value of $\rho_\chi  \simeq  0.4~\mathrm{GeV/cm}^3$.
We introduce the function $\Omega^-(r)$ as the interaction rate between dark matter and the target in the NS. This interaction rate corresponds to the case when the DM is captured and depends on factors such as the DM-target cross section and the target number density. Additionally, the function $\eta(r)$ takes into account the opacity of the star.
After the capture and annihilation of DM, and assuming that both processes are in equilibrium, we obtain the following heating luminosity observed at the distance,
 \begin{equation}
   L_\chi^{\infty} =  B(R_{\rm NS})  \times  \left[ X + (\gamma -1) \right]m_\chi C(m_\chi)  ~,
   \label{eq:luminosity_DM}
 \end{equation}
where $\gamma = B(R_{\rm NS}) ^{-1/2}$ is the Lorentz factor of the incoming DM particle. 
 We introduce $X$ to express the fraction of the annihilation energy transferred to the heat of the NS, and the $(\gamma-1)$ factor accounts for the contribution of the DM kinetic energy~\cite{Baryakhtar:2017dbj}.

Assuming the heating luminosity is dominated by the DM effects expressed in Eq.~\eqref{eq:luminosity_DM}, we can read out the surface temperature of NSs in this scenario from Eq.~\eqref{eq:L_heat=L_gamma}:
\begin{equation}
     T_{\rm  s}|_{\rm  DM}  \simeq  \left(\frac{[X+(\gamma-1)]m_\chi\, C(m_\chi)}{4\pi R_{\rm NS}^2\sigma_{\rm SB}}\right)^{1/4}.
     \label{eq:T_s-DM}
 \end{equation}
The maximum value of $T_{\rm  s}$ predicted, due to the heating by the capture of DM and its annihilation within a NS core, is $T_{\rm  s} \simeq 2600\,\rm K$~\cite{Chatterjee:2022dhp}. 
It is worth mentioning that the DM mass dependence in Eq.~\eqref{eq:T_s-DM} is cancelled by $C(m_\chi)  \propto  m_{\chi}^{-1}$. 
Therefore, the predicted surface temperature dominated by DM effects will be quite universal for a wide range of DM mass, $1~\mathrm{GeV}  \lesssim  m_{\chi}  \lesssim  1~\mathrm{PeV}$, if DM has a larger cross section with targets than the threshold value.

\section{Vortex creep heating}
\label{sec:Vortex_creep_heating}

The key to understanding the vortex creep heating is the neutron $^1S_0$ superfluid that is expected to appear in the inner crust of a NS~\cite{Page:2013hxa, Haskell:2017lkl, Sedrakian:2018ydt}. 
Except for this superfluid, the components, such as a rigid crust, a lattice of nuclei, charged particles, and NS core (denoted as \textit{crust component}), are tightly coupled through electromagnetic force and are directly affected by the external torque.
On the other hand, the neutron $^1S_0$ superfluid (denoted as \textit{superfluid component}) is only indirectly affected through the interaction with the crust component.  
This two-component description is originally proposed in the context of NS glitch phenomena~\cite{1969Natur.224..872B}. 
Under this description, we need to follow two angular velocities, $\bm{\Omega}_{\rm  c}$ and $\bm{\Omega}_{\rm  s}$ for the crust and superfluid components, respectively.
The friction between these two components is crucial for the vortex creep heating~\cite{1984ApJ...276..325A} as we will review below.

The interaction between the crust and superfluid components is mediated by the \textit{vortex line} of the neutron superfluid~\cite{Anderson:1975zze}. 
This string-shaped configuration appears in a rotating superfluid system and has a nonzero vorticity quantized in units of $\kappa \equiv h/(2m_n)$, where $h$ is the Planck constant and $m_n$ is the neutron mass. In a NS, the number of vortex lines saturates at the point where its integrated vorticity corresponds to the rigid body rotation of the superfluid component. 
In this saturation limit, the angular velocity of the superfluid component at a certain position decreases only if the number of vortex lines inside this position reduces. This is accomplished by the radial motion of vortex lines, known as  \textit{vortex creep}. See Ref.~\cite{Fujiwara:2023tmr} for more detailed discussions on the vortex creep.

The vortex creep occurs as a consequence of two forces exerting on vortex lines, \textit{pinning force} and \textit{the Magnus force}:
The pinning force, $\bm{f}_{\rm  pin}$, originates from the interaction between the vortex and the lattice nuclei and fixes a vortex line at some stable points in the crust. As a result, the angular velocity of superfluid remains constant, while that of the crust component slows down, leading to ${\Omega}_{\rm  s}  -  {\Omega}_{\rm  c} >0$.
This non-zero relative angular velocity will be a trigger for the Magnus force as expressed below:
\begin{align}
  \bm{f}_{\rm  Mag}  &=  \rho  \left[ ( \bm{\Omega}_{\rm  s}  -  \bm{\Omega}_{\rm  c})  \times  \bm{r} \right]  \times  \bm{\kappa},
  \label{eq:fmag}
\end{align}
where $\rho$ is the mass density of superfluid, and $\bm{r}$ denotes the position vector from the NS center. We introduce the vorticity vector denoted as $\bm{\kappa}$, which has the absolute value $|\bm{\kappa}| = \kappa$ and is aligned to the rotational axis with right-handed screw direction. 
Equation~\eqref{eq:fmag} shows that the direction of the Magnus force, and thus the direction of the vortex creep motion, is also fixed radially outward.

Depending on the NS temperature, vortex lines come off the pinning positions due to the thermal excitation or quantum tunnelling. 
According to the previous estimation~\cite{1991ApJ...373..592L,1993ApJ...403..285L,Larson:1998it}, the quantum tunnelling process is dominant for old NSs (with $T_{\rm  s}  \lesssim  10^6~\mathrm{K}$), and the unpinning rate for vortex lines is large enough for the system to be entered to the steady phase where the superfluid and the crust component have the same spin-down rate, $\dot{\Omega}_{\rm  c}  =  \dot{\Omega}_{\rm  s}  \equiv  \dot{\Omega}_{\infty}$. 
The relative angular velocity at this steady motion is nearly the same as that for the critical value where we have $f_{\rm  pin}  \simeq  f_{\rm  Mag}$. Using Eq.~\eqref{eq:fmag}, we obtain 
\begin{align}
  ( \Omega_{\rm  s}  -  \Omega_c )_\infty  
  \simeq  
  \frac{f_{\rm  pin}}{\rho  \kappa  r},
\end{align}
where $r$ is the distance from the rotational axis. 

The rotational energy stored in the neutron superfluid, which rotates faster than the crust component, is dissipated as heat via friction caused by the vortex creep motion. The energy dissipation due to this steady vortex creep motion gives the heating luminosity for NSs~\cite{1984ApJ...276..325A},
\begin{align}
  L_{\rm  VCH}^\infty  \simeq  B  ( R_{\rm  NS} )  \times  J  | \dot{\Omega}_\infty |,
  \label{eq:lvch}
\end{align}
where the proportional constant $J$ is defined as
\begin{align}
  J  \equiv  \int  d  I_{\rm  pin}  ( \Omega_{\rm  s}  -  \Omega_c )_\infty  \simeq \int   d  I_{\rm  pin}  \frac{f_{\rm  pin}}{\rho  \kappa  r}.
\end{align}
Here, $I_{\rm  pin}$ denotes the moment of inertia for the NS crust where the pinning force is relevant. 
The late-time surface temperature is given from Eq.~\eqref{eq:L_heat=L_gamma} once we assume the vortex creep heating dominates in a NS, 
\begin{align}
  T_{\rm  s}  |_{\rm  VCH}  \simeq  \left( \frac{J  |\dot{\Omega}_{\infty}|}{4  \pi  R_{\rm  NS}^2  \sigma_{\rm  SB}} \right)^\frac{1}{4}.
  \label{eq:ts_vch}
\end{align}
Note that the $J$ is only one parameter to characterize the predicted temperature for each NS since we can input $|\dot{\Omega}_\infty|$ and $R_{\mathrm{NS}}$ from observation.

The parameter $J$ is, in principle, determined by the pinning force that is expected to be universal for all NSs. This pinning force can be evaluated using a nuclear many-body calculation~\cite{1988ApJ...328..680E,Pizzochero:1997iq,Donati:2003zz,Donati:2006mfa,Donati:2004gnw,Avogadro:2006ed,Avogadro:2007kxn,Avogadro:2008uy,Seveso:2016,Klausner:2023ggr}, though the present theoretical calculation suffers from large uncertainty. In Ref.~\cite{Fujiwara:2023tmr}, we estimated the parameter $J$ using the semi-classical~\cite{Seveso:2016} and quantum~\cite{Klausner:2023ggr} calculations of the pinning force and obtain the following range of $J$:\footnote{We obtain larger values of $J$ if we use the pinning force computed in Refs.~\cite{Donati:2004gnw, Avogadro:2008uy} (see Ref.~\cite{Fujiwara:2023tmr}).} 
%
\begin{flalign}
  &J \! \in \! [ 3.9 \! \times  \! 10^{40},  1.9  \!\times \! 10^{43}  ]\,\mathrm{erg  \cdot  s} \quad \text{(Semi-classical~\cite{Seveso:2016})}~,
  \label{eq:J_Semi-classical}
  \\
 & J \! \in\!  [ 1.7 \! \times  \! 10^{40},  2.7 \! \times \!  10^{42} ]\,\mathrm{erg  \cdot  s} \quad  \text{(Quantum \cite{Klausner:2023ggr})} ~.
  \label{eq:J_Quantum}
\end{flalign}

On the other hand, we can also estimate the parameter $J$ from the temperature observation of old NSs through Eq.~\eqref{eq:ts_vch}. If the vortex creep heating dominates the heating luminosity, we expect that the values of $J$ estimated from Eq.~\eqref{eq:ts_vch} are almost universal over the NSs within the uncertainty coming from the surface temperature observation (typically a factor of a few error in $T_{\rm  s}$~\cite{Potekhin:2020ttj}) and from the NS structure (an $\mathcal{O}(1)$ factor in $J$ and $R_{\mathrm{NS}}$). In Ref.~\cite{Fujiwara:2023tmr}, we showed that the current temperature observations of old NSs indeed support this prediction, with $J$ in the range  
\begin{equation}
  J \simeq 10^{42.9 \text{--} 43.8}~\mathrm{erg} \cdot \mathrm{s} \quad 
  \text{(Observation~\cite{Fujiwara:2023tmr})}~. 
  \label{eq:J_obs}
\end{equation}
It is found that this observationally favoured range of $J$ can be consistent with the theoretical estimate in Eq.~\eqref{eq:J_Semi-classical}, though much smaller values are also allowed in the theoretical estimations.

\section{Results}
\label{sec:Results}
  \begin{figure}
  \centering
\includegraphics[width=0.483\textwidth]{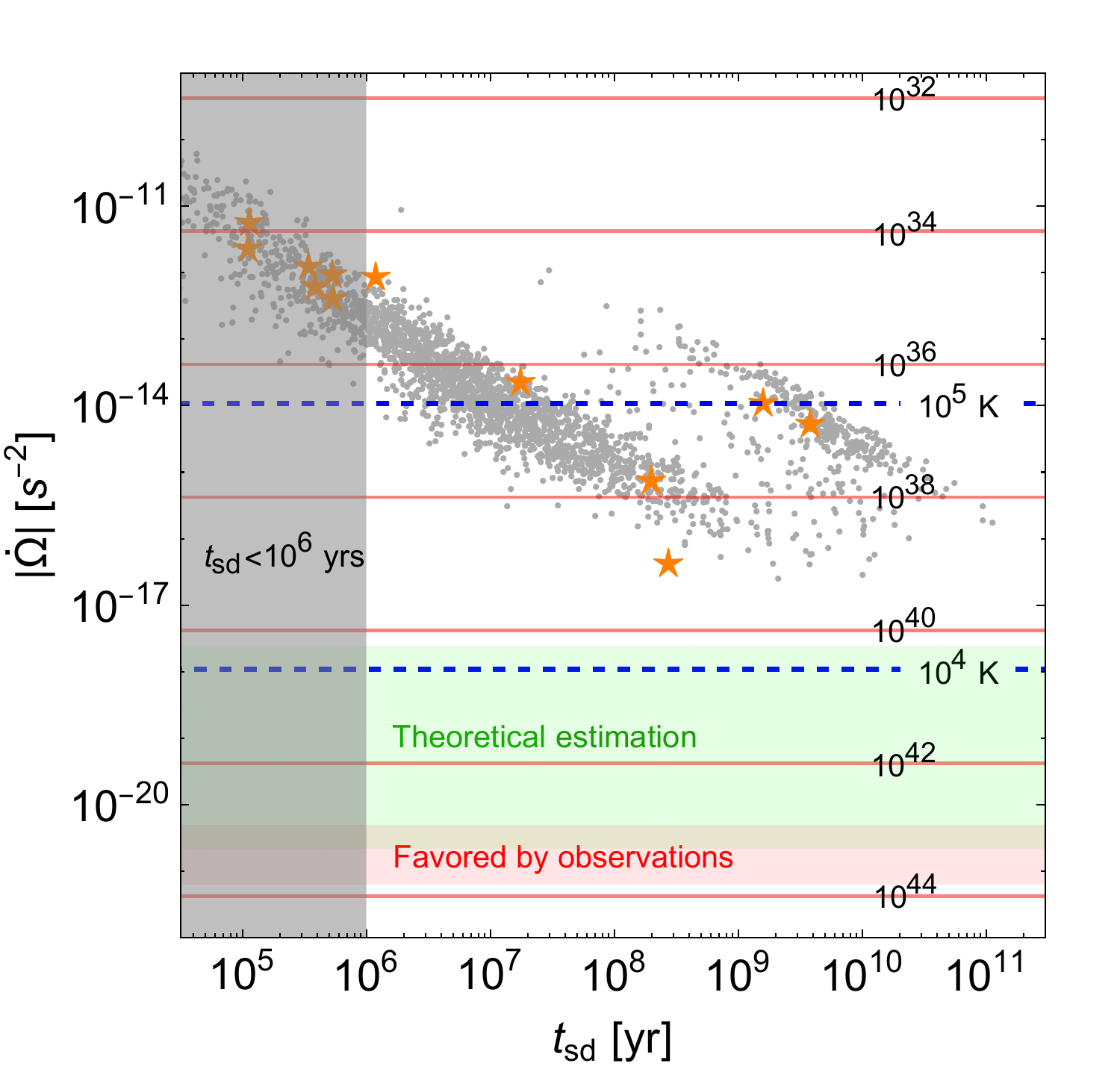}
\caption{Contours of $J$ $[\mathrm{erg} \cdot \mathrm{s}]$ corresponding to $L_{\mathrm{VCH}}^\infty = L_\chi^{\infty}$ in the $t_{\rm sd}$-$\dot{|{\Omega}|}$ plane (solid red lines). Grey dots represent spin-down pulsar data from the ATNF pulsar catalogue~\cite{Manchester:2004bp, ATNF_catalogue}, and orange stars indicate NSs used to test the quasi-universality of $J$ in vortex creep heating in Ref.~\cite{Fujiwara:2023tmr}. Green and red bands indicate the theoretical estimates~\eqref{eq:J_Semi-classical}, \eqref{eq:J_Quantum} and the observationally favored range~\eqref{eq:J_obs} of $J$, respectively. 
Blue dashed contours show the temperatures predicted by the vortex creep heating for $J=10^{42.9}~\mathrm{erg} \cdot \mathrm{s}$.
}
\label{fig:tsd_Omegadot}
 \end{figure}

Now we discuss the implications of the vortex creep heating mechanism for the prospects of the WIMP DM search via the NS surface temperature observation. For the DM heating to have a visible effect on the evolution of the NS surface temperature, its luminosity must dominate the heating luminosity. This requires $L_{\mathrm{VCH}}^\infty < L_\chi^{\infty}$, imposing an upper bound on $J$ in Eq.~\eqref{eq:lvch}. In Fig.~\ref{fig:tsd_Omegadot}, we show contours of $J$ in units of $\mathrm{erg} \cdot \mathrm{s}$ corresponding to $L_{\mathrm{VCH}}^\infty = L_\chi^{\infty}$ in the $t_{\rm sd}$-$\dot{|{\Omega}|}$ plane (red solid lines), where $t_{\rm sd}$ is the spin-down age of the NS; in this evaluation, {we take $T_{\rm  s}  = 
 2600~\mathrm{K}$ in Eq.~\eqref{eq:L_heat=L_gamma} to estimate the most optimistic value of the DM heating luminosity as estimated in Ref.~\cite{Chatterjee:2022dhp}}.
We also show the spin-down pulsar data taken from the Australian Telescope National Facility (ATNF) pulsar catalogue~\cite{Manchester:2004bp, ATNF_catalogue} as the grey dots. The orange stars indicate the NSs used to test the vortex creep heating in Ref.~\cite{Fujiwara:2023tmr}. 
The grey-shaded region corresponds to NSs too young to probe DM heating ($t_{\rm sd}<10^6~\mathrm{yrs}$). 
As we see, for the observed values of $|\dot{\Omega}|$, we need $J \lesssim \mathcal{O} (10^{36-39})~\mathrm{erg} \cdot \mathrm{s}$ in order for DM heating to prevail over vortex heating. These values are smaller than both the theoretical estimations~\eqref{eq:J_Semi-classical}, \eqref{eq:J_Quantum} (green band) and the observationally favoured range~\eqref{eq:J_obs} (red band) by orders of magnitude. 

In other words, if the value of $J$ is in the observationally favoured range, $J  \sim  10^{43}  \text{--}  10^{44}~\mathrm{erg  \cdot  s}$, the surface temperature of NSs becomes much higher than the maximal value predicted by the DM heating, $T_{\rm  s}  \simeq  2600~\mathrm{K}$~\cite{Chatterjee:2022dhp}. To see this, we also show the surface temperatures predicted by the vortex creep heating for $J=10^{42.9}~\mathrm{erg} \cdot \mathrm{s}$ {(the lower value in observationally favoured range)} in blue dashed contours in Fig.~\ref{fig:tsd_Omegadot}. We found that for all of the known pulsars $T_{\rm  s} \gg 2600$~K, implying that the vortex heating effect always conceals the DM heating effect.

\begin{figure}
  \centering
\includegraphics[width=0.483\textwidth]{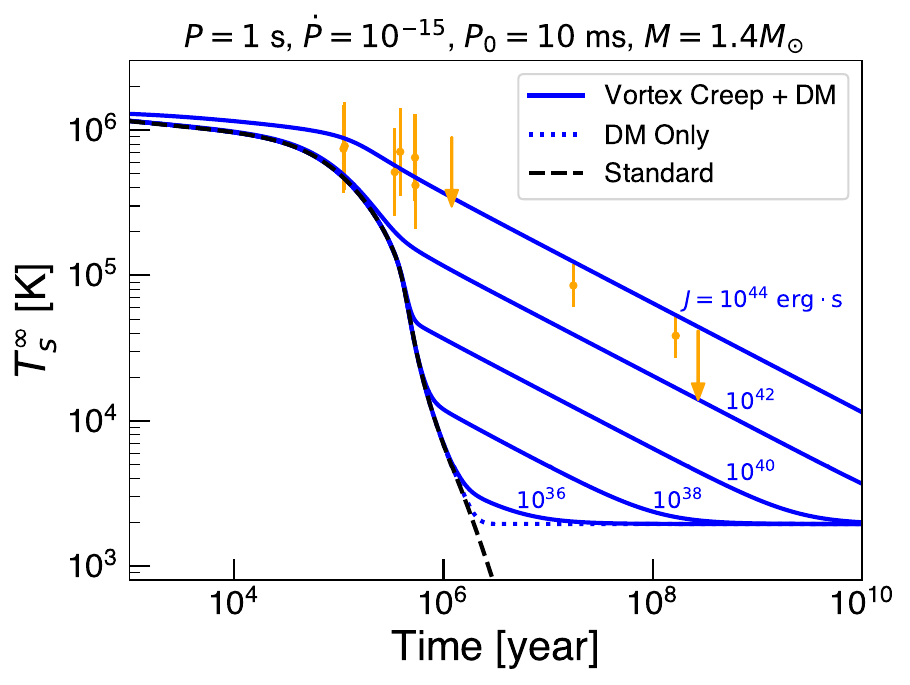}
\caption{Time evolution of the surface temperature in the presence of both the vortex creep and DM heating in the blue solid lines, where we set $P\dot{P} = 10^{-15}$~s, $P_0 = 10~\mathrm{ms}$, $\rho_\chi = 0.42~\mathrm{GeV} \cdot \mathrm{cm}^{-3}$, $v_d = 230~\mathrm{km} \cdot \mathrm{s}^{-1}$, $v_\star = 0$, and $X=1$. The blue dotted and black dashed lines show the cases for only the DM heating and the standard cooling, respectively. The orange dots represent the temperature data for ordinary pulsars considered in Ref.~\cite{Fujiwara:2023tmr}, where the bars show the error in $T_{\rm  s}$ and the arrows indicate the upper limits.  }
\label{fig:teff_o}
\end{figure}

In Fig.~\ref{fig:teff_o}, we show the time evolution of the surface temperature in the presence of both the vortex creep and DM heating in the blue solid lines.\footnote{We use the the Akmal-Pandharipande-Ravenhall (APR)~\cite{Akmal:1998cf} equation of state for a NS mass of $1.4M_\odot$ to calculate the NS structure. For Cooper pairing gap models, we use the SFB model~\cite{Schwenk:2002fq} for the neutron singlet pairing, the model ``b'' in Ref.~\cite{Page:2004fy} for the neutron triplet pairing, and the CCDK model~\cite{Chen:1993bam} for the proton singlet pairing. The late-time temperatures rarely depend on the choice of these models.} We take $J = 10^{44}$, $10^{42}$, $10^{40}$, $10^{38}$, and $10^{36}~\mathrm{erg} \cdot \mathrm{s}$ from top to bottom. In the calculation of the vortex creep heating, we assume that the magnetic dipole radiation causes the pulsar slowdown, and we set $P\dot{P} = 10^{-15}$~s and $P_0 = 10~\mathrm{ms}$, where $P$, $\dot{P}$, and $P_0$ are the period, its time derivative, and initial period of the pulsar, respectively. This choice corresponds to the surface magnetic field of $B_s \simeq 10^{12}$~G, which is a typical value for ordinary pulsars.\footnote{In millisecond pulsars, it is likely that another heating mechanism called the rotochemical heating~\cite{Reisenegger:1994be, 1992A&A...262..131H, 1993A&A...271..187G, Fernandez:2005cg, Villain:2005ns, Petrovich:2009yh, Pi:2009eq, Gonzalez-Jimenez:2014iia, Yanagi:2019vrr} (see also Refs.~\cite{Gusakov:2015kaa, Kantor:2021vwj}) operates and its effect will hide the DM heating effect~\cite{Hamaguchi:2019oev}. For this reason, we focus on ordinary pulsars in the present analysis.  } In the calculation of the DM heating, we use $\rho_\chi = 0.42~\mathrm{GeV} \cdot \mathrm{cm}^{-3}$~\cite{Pato:2015dua}, $v_d = 230~\mathrm{km} \cdot \mathrm{s}^{-1}$, $v_\star = 0$, and $X=1$, for which we have $T_{\rm  s}^{\infty}|_{\rm  DM} \simeq 2 \times 10^3$~K. The blue dotted and black dashed lines show the cases for only the DM heating and the standard cooling, respectively. The orange dots represent the temperature data for ordinary pulsars considered in Ref.~\cite{Fujiwara:2023tmr}, where the bars show the error in $T_{\rm  s}^{\infty}$ and the arrows indicate the upper limits. This figure shows that for observationally favoured values of $J$ in Eq.~\eqref{eq:J_obs}, the late-time temperature evolution is governed by the vortex creep heating, and the effect of the DM heating is completely invisible. The DM heating effect can be comparable to the vortex creep heating effect only if the value of $J$ is smaller than those values by orders of magnitude; for $J = 10^{36}~\mathrm{erg} \cdot \mathrm{s}$ ($10^{38}~\mathrm{erg} \cdot \mathrm{s}$), the DM heating may be observable for NSs older than $t \gtrsim 10^7$~years ($3 \times 10^8$~years).

\section{Conclusions \& Discussion}
\label{sec:Conclusions}

We have studied the effect of the vortex creep heating on the search of WIMP DM through the surface temperature observation of old NSs. We have found that for values of $J$ that are favoured by both theory (Eqs.~\eqref{eq:J_Semi-classical} and \eqref{eq:J_Quantum}) and observation (Eq.~\eqref{eq:J_obs}), the vortex creep heating effect is much stronger than the DM heating effect, and thus it is challenging to detect the signature of WIMP DM through the temperature observations of old NSs. For this to be possible, $J$ should be smaller than these values by orders of magnitude, as small as $J \lesssim 10^{38}~\mathrm{erg} \cdot \mathrm{s}$. 

It is, of course, possible that future computations of the vortex-nuclei interaction will indicate a much smaller pinning force than the present calculations, leading to a sufficiently small value of $J$. Another possibility to suppress the vortex creep heating is that the vortex creep rate is much lower than the theoretical estimates~\cite{1991ApJ...373..592L,1993ApJ...403..285L,Larson:1998it} so that the spin-down rate of the superfluid component does not follow that of the crust component. In this case, the spin-down of the superfluid component does not reach the steady phase, which is assumed in the derivation of Eq.~\eqref{eq:lvch}, and the vortex heating luminosity depends on $|\dot{\Omega}_{\mathrm{s}}|$. It is found that $\dot{\Omega}_{\mathrm{s}}$ is proportional to the vortex creep rate (see Ref.~\cite{Fujiwara:2023tmr}), and therefore a very small creep rate could suppress the vertex heating. To explore these possibilities, improvements in the theoretical calculations of the vortex-nuclear interactions are highly motivated.  

Besides, it is also important to accumulate more data of the surface temperatures of old NSs to observationally test the vortex creep heating mechanism as discussed in Ref.~\cite{Fujiwara:2023tmr}. We envision that this will be performed in the near future through improved optical, UV, and X-ray observations.

\begin{acknowledgments}
This work is supported in part by the Collaborative Research Center SFB1258 and by the Deutsche Forschungsgemeinschaft (DFG, German Research Foundation) under Germany's Excellence Strategy - EXC-2094 - 390783311[MF], 
JSPS Core-to-Core Program (No.JPJSCCA20200002 [MF]), the Grant-in-Aid for Innovative Areas (No.19H05810 [KH and MRQ], No.19H05802 [KH], No.18H05542 [MF and NN]), Scientific Research B (No.20H01897 [KH and NN]), Young Scientists (No.21K13916 [NN]).
\end{acknowledgments}

\bibliographystyle{name}
\bibliography{references}

\end{document}